\documentclass[prd,onecolumn,notitlepage,showpacs,preprintnumbers,amsmath,amssymb,nofootinbib,APS,10pt,superscriptaddress]{revtex4-2}

\usepackage{dcolumn}
\usepackage{bm}
\usepackage[spanish,english]{babel}
\usepackage[utf8]{inputenc}
\usepackage{amsmath,amssymb,amsfonts,latexsym,cancel}
\usepackage{graphicx}
\usepackage{color}
\usepackage{soul}
\usepackage[normalem]{ulem}
\usepackage{hyperref}
\usepackage{slashed}
\usepackage{float}
\usepackage{xcolor}
\usepackage[shortlabels]{enumitem}

\allowdisplaybreaks

\newcommand{\be}{\begin{equation}}
\newcommand{\ee}{\end{equation}}
\newcommand{\bea}{\begin{eqnarray}}
\newcommand{\eea}{\end{eqnarray}}

\begin{document}

\title{Low energy states and $CPT$ invariance at the big bang}

\author{Sergi Nadal-Gisbert}\email{sergi.nadal@uv.es}
\affiliation{Departamento de Fisica Teorica and IFIC, Centro Mixto Universidad de Valencia-CSIC. Facultad de Fisica, Universidad de Valencia, Burjassot-46100, Valencia, Spain.}
\affiliation{Department of Physics and Astronomy, Louisiana State University, Baton Rouge, Louisiana 70803, U.S.A.}
\author{Jos\'e Navarro-Salas}\email{jnavarro@ific.uv.es}

\affiliation{Departamento de Fisica Teorica and IFIC, Centro Mixto Universidad de Valencia-CSIC. Facultad de Fisica, Universidad de Valencia, Burjassot-46100, Valencia, Spain.}

\author{Silvia Pla}\email{silvia.pla\_garcia@kcl.ac.uk}
\affiliation{Theoretical Particle Physics and Cosmology, King’s College London, WC2R 2LS London, United Kingdom.}
%\affiliation{Departamento de Fisica Teorica and IFIC, Centro Mixto Universidad de Valencia-CSIC. Facultad de Fisica, Universidad de Valencia, Burjassot-46100, Valencia, Spain.}
%\affiliation{Consortium for Fundamental Physics, School of Mathematics and Statistics,
%Hicks Building, Hounsfield Road, Sheffield. S3 7RH United Kingdom. }
%

%

\begin{abstract}

In this paper, we analyze the quantum vacuum in a radiation-dominated and $CPT$-invariant universe by further imposing the quantum states to be ultraviolet regular i.e., satisfying the Hadamard/adiabatic condition. For scalar fields, this is enforced by constructing the vacuum via the States of Low Energy proposal. For   
 spin-$\frac{1}{2}$ fields, we extend this proposal for a FLRW spacetime and apply it for the radiation-dominated and $CPT$-invariant universe. We focus on minimizing the smeared energy density around the big bang and give strong evidence that the resulting states satisfy the Hadamard/adiabatic condition. These states are then self-consistent candidates as effective big bang quantum vacuum from the field theory perspective.

\end{abstract}

\date{\today}
\maketitle

\tableofcontents

\section{Introduction}

One of the basic issues in the  %traditional approach to the 
theory of quantized fields in curved spacetime \cite{birrell-davies, Fulling,  parker-toms, Hu} 
is the fixing of a preferred vacuum state. The canonical quantization approach soon exhibited that the vacuum state is not unique in a time-dependent spacetime. %\cite{Parker66,Parker68,Parker69,Parker71}. 
Only for stationary spacetimes, or adiabatic regions in expanding universes, %approximately stationary (i.e., adiabatic regions), 
one can naturally find a privileged definition of the vacuum.  A major consequence of this ambiguity is the particle creation phenomena, as first discovered in cosmology \cite{Parker66,Parker68,Parker69,Parker71} and also in the vicinity of black holes \cite{Hawking}. 
However, a requirement that a physically admissible quantum state should satisfy is the Hadamard condition \cite{Wald94}, which specifies the singularity structure of the two-point function. This condition ensures the existence of Wick polynomials of arbitrary order \cite{BF2000, hollands-wald-01, hollands-wald-02} and hence, the perturbative series of an interacting theory to be well-defined at any order. The necessity of the Hadamard condition was further motivated in \cite{Fewster13}. The  Hadamard condition, which is defined for arbitrary spacetimes, can be transformed into the adiabatic condition \cite{Fulling, parker-toms, Pirk93, Hollands01} in homogeneous spacetimes. The  adiabatic condition  fixes the large momentum structure of admissible states.  \\
%is a natural  generalization to arbitrary spacetimes of the adiabatic condition , which is only defined in homogeneous spacetimes and determines the large momentum structure of admissible states. \\

%$$ HADAMARD > DeWitt-Schwinger \equiv Adiabatic \ \ regularization $$

%A second difficulty is to fix  the initial vacuum state. 
%In general, 

On a generic time-varying spacetime one cannot single out a preferred vacuum. This lack of a unique vacuum choice is strongly manifested in Friedmann-Lemaître-Robertson-Walker (FLRW) spacetimes.  Several approaches have been considered to select a preferred vacuum state at early times, based on different viewpoints \cite{Grib:1976pw, Mamaev, ChitreHartle, Fulling79, 4adiabatic, agullo-nelson-ashtekar}, from which one can predict late time quantum effects.   Furthermore, it is also highly nontrivial to obtain a vacuum state satisfying the Hadamard condition. An especially appealing proposal %, inspired by the work \cite{Fewster2000}, 
is the  States of Low Energy (SLE) prescription \cite{Olbermann07, Banerjee20}. These states of low energy are obtained by minimizing  the expectation value of the energy density after smearing it with a time-dependent  test function. This prescription ensures the Hadamard condition or, equivalently, the (all orders) adiabatic condition \cite{parker-toms}. The SLE were introduced originally only for scalar fields (and for minimal coupling). It is easy to show that the Minkowski vacuum is the state of low energy, irrespectively of the specific form of the smearing function. For simple asymptotically flat regions, where the expansion factor $a(t)$ approaches constant values at $t\to \pm \infty$, the initial and final Minkowski vacua, corresponds to the state of minimal energy when the time averaging has support at early and late times respectively. For de Sitter spacetime,  the Bunch-Davies vacuum is the state of low energy, irrespective of the particular form of the smearing function, as long as it has support at the distant past \cite{Degner2010,Olmedo}. However,  the  states of low energy depend, in general,  on the choice of the smearing function.  This method was recently applied to obtain physically motivated vacua in the Schwinger effect \cite{Alvarez-Dominguez:2023ten} and for scalar fields with Yukawa interaction \cite{Ferreiro:2023xer}.\\

Of special physical interest is to analyze this issue for a radiation-dominated universe. The study of this particular expansion rate has different motivations. It can be thought of as a natural pre-inflationary phase, as it has been recently discussed in \cite{AndersonCarlson20, Das:2014ffa, Ordines2022}.  The vacuum in this pre-de Sitter space should smoothly evolve to a state that is approximately equivalent to the Bunch-Davies state for large $k$'s, but differs significantly from it at small $k$'s. Furthermore, from a noninflationary perspective, the issue of how to define a preferred vacuum in a radiation-dominated universe is also of special relevance in general, particularly in relation to the interesting and recent proposal of Ref. \cite{LetterCPT, Large}. In this view of cosmic evolution, the CPT symmetry plays a crucial role to single out privileged vacua. The gravitational background is assumed to be time-reversal symmetric with respect to the big bang event $\tau=0$. A radiation-dominated era emerges from this special event, for which a natural analytic continuation of the  expansion factor $a(\tau) \sim \tau$ to negative values of conformal time $\tau$ is assumed.   A radiation-dominated universe going back to the big bang is also the simplest way to enforce Penrose's  Weyl curvature hypothesis \cite{Penrose79}.
\\

The goals of this paper are:
\begin{enumerate}[{ i)}]

 \item Extend the prescription of states of low energy  for spin-$\frac{1}{2}$ fields  on FLRW spacetimes.
    \item Characterize the possible $CPT$-invariant Hadamard states for scalars and spin-$\frac{1}{2}$ fields just at the big bang event $\tau=0$, obtaining a  $\left\{\theta_k, \Theta_k\right\}$-family of $CPT$-invariant Hadamard states for scalars and fermions respectively.
    
    \item Study the possible $CPT$-invariant States of Low Energy in a radiation-dominated universe and analyze whether the smeared energy density can be minimized around the big bang for both scalar and spin-$\tfrac{1}{2}$ fields. We check that the resulting family of $CPT$-invariant states of low energy, indeed satisfy the Hadamard/adiabatic condition. %stated for $\left\{\theta_k, \Theta_k\right\}$ .
\end{enumerate}

  For pedagogical reasons the organization of the paper is as follows.  %organization of the paper follows the route. Accordingly, the paper is organized as follows. 
 In Sec. \ref{sec:rad-dom-CPT-scalars} we parametrize the possible $CPT$-invariant states in a radiation-dominated universe for scalar fields and give the asymptotic ultraviolet condition that they have to satisfy in order to be  Hadamard. In Sec. \ref{sec::SLE::Scalars} we review the SLE characterization, extending it for a general coupling to the scalar curvature and paying special attention to the big bang singularity. We restrict the SLE characterization to the $CPT$-invariant states. In Sec. \ref{Subsec::BigBang::Scalars}, 
we study the possible SLE with smearing functions with support around the big bang for scalar fields. In Sec. \ref{sec:spinor-CPT} we parametrize the possible $CPT$-invariant states in a radiation-dominated universe for spin-$\frac12$ fields and characterize the asymptotic ultraviolet condition that they have to satisfy in order to be  Hadamard. In Sec. \ref{Sec::SLE::Fermions} we provide the SLE characterization for fermions.  In Sec. \ref{Sec::SLE::Fer::Early}, 
we study the possible SLE with smearing functions with support around the big bang for spin-$\frac12$ fields. Finally, In Sec. \ref{sec:conclusions} we summarize our results and discussions. Most of the computations in this paper have been done with the aid of the \emph{Mathematica} software. Throughout this paper, we use units in which ${\hbar=c=1}$.%\footnote{ We remark that the content of this paper improves and extends the analysis made in \cite{Beltran-Palau:2022vfg}. Here we focus on quantum states satisfying the Hadamard condition (or the adiabatic condition to all orders), while in \cite{Beltran-Palau:2022vfg} the ultraviolet requirement was to ensure only the renormalizability of the stress-energy tensor.} 

\section{$CPT$-invariant states for scalars in a radiation-dominated spacetime}
\label{sec:rad-dom-CPT-scalars}

In this section, we consider a massive scalar field $\phi$ propagating in a flat FLRW spacetime
\be \text{d}s^2 = a^2(\tau)(\text{d}\tau^2 - \text{d} \vec x^2) \ . \ee
 We will assume a radiation-dominated universe. The expansion factor is given, in conformal time, by
 \be a(\tau) \propto \tau \ . \ee 
It is convenient to expand the quantized field in Fourier modes adapted to the underlying homogeneity of the $3$-space
\be \label{modecomp}
    \phi(\tau,\vec x)= \int \frac{\text{d}^3 k}{\sqrt{2(2\pi)^{3}}}\left(A_{\vec k} e^{i\vec k \vec x} \phi_k(\tau) + A_{\vec k}^{\dagger} e^{-i\vec k \vec x} \phi_k^{*}(\tau)\right) \ ,
\ee
where the creation and annihilation operators satisfy the usual commutation relations ($[A_{\vec k}, A^\dagger_{\vec k'}] = \delta^3(\vec k - \vec k')$, etc). The normalization of the modes is fixed by the condition
\be \phi_k\phi'{}^{*}_k - \phi_k' \phi_k^* = \frac{2i}{a^2} \ , \ee
where the prime denotes derivative with respect to the conformal time. 
For our purposes it is convenient  to work with the rescaled Weyl field $\varphi \equiv a \,\phi$ and the rescaled  modes $\varphi_k \equiv a\,\phi_k$. The field equation implies ($a^2m^2= \gamma^2 \tau^2$)
\be\label{eqm}
  \varphi_k''(\tau ) + \left[k^2+\gamma^2 \tau^2 \right] \varphi_k(\tau)=0 \ , 
\ee
and the normalization condition is given by
\be \label{Ncondition2}\varphi_k  \varphi'{}^{*}_k - \varphi_k' \varphi_k^* = 2i \ . \ee

The general solution of \eqref{eqm} can be expressed in terms of parabolic cylindrical functions $D_{\nu}(z)$ \cite{booksf}
\be\label{sol-scalars}
\varphi_k(\tau)= C_{k,1} \,S_k(\tau) +C_{k,2}\,  S_k^*(-\tau)\, ,
\ee
where
\be
S_k(\tau)=\frac{1}{(2 \gamma)^{1/4}}\, D_{-\frac{1}{2}-2i\kappa }\big(e^{i\tfrac{\pi}{4}}
  \sqrt {2\gamma } \,\tau\big)\, ,
\ee
and \be \label{eq:kappa:k} \kappa=\frac{k^2}{4 \gamma}\, .\ee Any choice of $k$-functions $C_{k,1}$ and $C_{k,2}$ defines a set of modes characterizing a given vacuum state.\footnote{From the Wronskian condition \eqref{Ncondition2} we get the following normalization condition for $C_{k,1}$ and $C_{k,2}$: $$ \frac{e^{ \pi \kappa}}{2}(|C_{k,1}|^2+|C_{k,2}|^2)+\frac{\sqrt{2}\cosh{(2 \pi \kappa)}}{\sqrt{\pi}}\textrm{Re}[e^{-i\frac{\pi}{4}}C_{k,1}C^*_{k,2}\Gamma(\tfrac{1}{2}-2i\kappa)]=1$$} All these vacua are, by construction, invariant under spatial translations, rotations, parity and charge conjugation (which is here trivial since our scalar field is real). Time translation is not a symmetry for an expanding universe, and time-reversal has not been considered so far.  As remarked in the introduction, there is no natural way to select a preferred Fock vacua.
In the radiation-dominated universe, and due to the very special form of the expansion factor in conformal time, one can further reduce the freedom in choosing a vacuum by exploiting the time-reversal symmetry  $\tau \to -\tau$ of the background $\text{d}s^2 \propto \tau^2(\text{d}\tau^2 -\text{d}\vec x^2$).  \\  

%Generically, the action of charge conjugation $C$, parity $P$, and time-reversal $T$ on a classical  scalar field is given by \cite{AGbook} $C:\phi(\tau,\vec{x}) \to \xi_c^*\phi^{*}(\tau,\vec{x}) $; $P:\phi(\tau, \vec x)\to \xi_p^*\phi (\tau, -\vec x)$; $T\phi(\tau, \vec x) \to \xi_t^* \phi^*(-\tau, \vec x)$. [The $\xi$'s are the associated phases of the $C$, $P$, $T$ transformations]. In the quantized theory, $C$ and $P$ are representated as unitary operators, while $T$ is converted into a antiunitary operator. In our case $C$ and $P$ are trivially implemented in the assumed Fourier expansion. Enforcing $T$ is the key ingredient. However,  it seems more useful \cite{LetterCPT, Large} to consider $CPT$ all at once in the analysis (furthermore, we also choose $\xi_c\xi_p\xi_t=1$ \cite{Weinberg} and $CPT \phi(x) (CPT)^{-1} = \phi(-x)^{\dagger}$).  The condition for a $CPT$-invariant vacuum state takes a very  simple form on the time-dependent part of  scalar field modes:  {\color{blue} $\varphi_k(-\tau) =  -\varphi_k^*(\tau)$, where the minus sign comes from the factor $a(\tau)$ in the definition of the rescaled field $\varphi$. This sign can be reabsorbed by redefining the modes $\varphi_k \to i\varphi_k$. Taking into account this phase redefinition we can write}

Following \cite{LetterCPT, Large}, we  directly study the effect of charge conjugation $C$, parity $P$, and time-reversal $T$ on the Weyl transformed scalar field $\varphi$. The action of these transformations is \cite{AGbook},  $C:\varphi(\tau,\vec{x}) \to \xi_c^*\varphi^{*}(\tau,\vec{x}) $; $P:\varphi(\tau, \vec x)\to \xi_p^*\varphi (\tau, -\vec x)$; $T:\varphi(\tau, \vec x) \to \xi_t^* \varphi^*(-\tau, \vec x)$. [The $\xi$'s are the associated phases of the $C$, $P$, $T$ transformations]. In the quantized theory, $C$ and $P$ are represented as unitary operators, while $T$ is converted into a antiunitary operator. In our case $C$ and $P$ are trivially implemented in the assumed Fourier expansion. Enforcing $T$ is the key ingredient. However,  it seems more useful to consider $CPT$ all at once in the analysis [furthermore, we also choose $\xi_c\xi_p\xi_t=1$ \cite{Weinberg}; therefore $CPT \varphi(x) (CPT)^{-1} = \varphi(-x)^{\dagger}$].\footnote{We note that if the $T$ transformation is directly applied on the original scalar field $\phi(x)$ there will be a difference of sign with respect to the transformation defined here. This sign can be always absorbed in the phase $\xi_t^*$ or  by making a redefinition of the (Weyl transformed) modes $\varphi_k \to i\varphi_k$ to keep our choices (e.g., $\xi_c\xi_p\xi_t=1$), and hence, the final form of the $CPT$ transformation unaltered. For convenience we have adopted the conventions used in \cite{LetterCPT, Large}.}  The condition for a $CPT$-invariant vacuum state takes a very  simple form on the time-dependent part of  scalar field modes,
\be \varphi_k(-\tau) =  \varphi_k^*(\tau) \ . \ee 
Using standard properties of the parabolic cylindrical functions it can be easily shown that in terms of the general solution \eqref{sol-scalars}, the condition above implies $C_{k,1}=C_{k,2}^*$.  It seems natural to characterize the $CPT$-invariant state by specifying initial data at $\tau=0$.  In the limit of $\tau\to 0$, the mode equation becomes
\begin{equation} \label{eq:massles-t0}
     \varphi_k''(\tau)+k^2\varphi_k(\tau)\sim 0\, .
\end{equation}
Then,
\be
\varphi_k(\tau)\sim c_k\, e^{-i k\tau}+ d_k\, e^{ i k\tau}\, .
\ee
The normalization condition  implies
$    |c_k|^2-|d_k|^2=k^{-1}$.
 This condition, together with the required $CPT$-invariance [%$c_k+d_k=c_k^{*}+d_k^{*}$ and $c_k-d_k=c_k^{*}-d_k^{*}$, or equivalently 
 $c_k=c^*_k$ and $d_k=d_k^*$, i.e., $c_k$ and $d_k$ must be real] allows us to reparametrize the constants $c_k$ and $d_k$ in terms of an hyperbolic angle $\theta_k$ as
\begin{equation} \label{eq:constantsCPT}
c_k=\frac{\cosh(\theta_k)}{\sqrt{ k}}\, , \quad d_k=\frac{\sinh(\theta_k)}{\sqrt{ k}}\,. 
\end{equation}
Therefore, the $CPT$-invariant solution for $\tau \to 0$ should go as 
\begin{equation} \label{eq:CPT-tau0}
    \varphi_k(\tau)\sim \frac{1}{\sqrt{ k}} e^{-ik\tau}\cosh{\theta_k}+ \frac{1}{\sqrt{k}} e^{ik\tau}\sinh{\theta_k}\, .
\end{equation}
At $\tau=0$ this results into 
\begin{equation} 
\varphi_k(0)= \frac{e^{\theta_k}}{\sqrt{ k}} \, , \quad \varphi'_k(0)= -i \sqrt{k}e^{-\theta_k} \, .
 \label{eqn:modesCPT}
\end{equation}
where $\theta_k$ is an arbitrary real function. In summary, the CPT requirement  reduces the space of possible vacuum states to a  family of states characterized by the hyperbolic initial ($\tau=0$) phase $\theta_k$. In terms of this parameter, the functions $C_{k,1}$ and $C_{k,2}$ read\footnote{The final expression for $C_{k,1}$ can be written in several ways by using some properties of the gamma functions. In particular we have used $\Gamma(z)\Gamma(z+\frac{1}{2})=2^{1-2z}\sqrt{\pi}\,\Gamma(2z)$ and $\Gamma(z)\Gamma(1-z)=\frac{\pi}{\sin(\pi z)}$.}
\be \label{eq:C1-CPT-scalars}
C_{k,1}=2^{i \kappa }\sqrt{\pi}e^{\pi \kappa}\Bigg(\frac{e^{-\frac{i \pi}{4}}e^{\theta_k}}{\kappa^{\frac{1}{4}}\Gamma(\frac{1}{4}-i\kappa)}+\frac{i\,\kappa^{\frac{1}{4}} e^{-\theta_k}}{\Gamma(\frac{3}{4}-i\kappa)}\Bigg)\, ,
\ee
  with $\kappa$  defined in \eqref{eq:kappa:k} and $C_{k,2}=C^*_{k,1}$. In other words, any $CPT$-invariant solution can be written in terms of the $\theta_k$ angle as
\be \label{eq:sol-scalars-cpt}
\varphi^{CPT}_k(\tau)=C_{k,1}\, S_k(\tau)+C_{k,1}^*\,S_k^*(-\tau)\, ,
\ee
with $C_{k,1}$ given above. We can regard this result as an equivalent characterization of the $CPT$-invariant vacua proposed in \cite{LetterCPT, Large}.  The main advantage of this reparametrization is that it allows us to characterize the nontrivial ultraviolet behavior of the modes at large $k$.  Also, in the Heisenberg picture we do understand now the time reversal vacuum state $|0 \rangle$ as defined by giving initial data $\varphi^{CPT}_k(\tau_0)$ at $\tau_0=0$. \\%

%\subsubsection{Late- and early-times vacua} \label{subsubsec:later-early-scalars}
To clarify the discussion above, it is interesting to present states that are not $CPT$-invariant. As time evolves the expansion of the Universe slows down and one can naturally define a late-times (infinite order) adiabatic vacuum $|0_+\rangle$ \cite{Birrell-Whitakker}.  In this asymptotic region ($\tau \to \infty$), the general solution \eqref{sol-scalars} behaves as a linear combination of positive and negative-frequency solutions, and the preferred (positive-frequency) solution for the late-times modes at $\tau\to \infty$ reads
\be\label{eq:adiabatic-late-scalars}
\varphi_k^{(+)}(\tau)\sim \frac{e^{-i\int_\tau \omega(u)du}}{\sqrt{\omega(\tau)}}\sim \frac{e^{-i \left( \frac{\gamma}{2}\tau^2+\kappa \ln(2 \gamma \tau^2 )\right)}}{\sqrt{\gamma \tau}}\, ,
\ee
where $\omega=\sqrt{k^2+m^2 a^2}$. The constants $C_{k,1}$ and $C_{k,2}$ can be fixed imposing this late-times behavior. We directly find
\be
C_{k,1}=\sqrt{2}\,  e^{-\frac{\pi \kappa}{2}} e^{i\frac{\pi}{8}}\, , \qquad C_{k,2}=0\, .
\ee
We note that the late-times adiabatic vacuum is not $CPT$-invariant, since $C_{k,1}\neq C_{k,2}^{*}$. Analogously, the early-times adiabatic vacuum $|0_{-}\rangle$ is determined by the asymptotic condition ($\tau \to- \infty$)
\be \label{eq:adiabatic-early-scalars}
\varphi_k^{(-)}(\tau)\sim \frac{e^{-i \int_\tau \omega (u ) \, du
   }}{\sqrt{\omega (\tau )}}\sim \frac{e^{i \left(\frac{\gamma}{2} \tau ^2+\kappa \log
   (2\gamma(-\tau)^2 )\right)}}{\sqrt{-\gamma  \tau
   }}\, ,
\ee
and $C_{k,1}=0$, $C_{k,2}=\sqrt{2}e^{-\frac{\pi \kappa}{2}}e^{-i\frac{\pi}{8}}$, which is also not $CPT$-invariant. Given the constants $C_{k,1}$ and $C_{k,2}$ for each solution, it is direct to see that $CPT |0_{\pm}\rangle=|0_{\mp}\rangle$.

\subsection{Ultraviolet regularity of the $CPT$-invariant vacuum states}\label{sec:UVreg-scalars}

For a quantum state  $|0\rangle$, to be admitted as physically acceptable we should demand it to be ultraviolet regular. This means that the high-energy behavior of the state must approach the behavior of Minkowski space at a rate such that basic composite operators can be renormalized. In cosmological backgrounds this translates into the {\it adiabatic condition}: for large $k$, the behavior of the field modes $\varphi_k$ should follow the Wentzel-Kramers-Brillouin (WKB) type asymptotic condition at all orders  %the Wentzel-Kramers-Brillouin (WKB) ansatz 
\cite{parker-fulling1,parker-fulling2,parker-fulling3, Anderson-Parker}
\be \label{conformalmodes}\varphi_k(\tau)\sim \frac{1}{\sqrt{\Omega_k(\tau)}}e^{-i\int^\tau \Omega_k(\tau') d\tau'}  \ , \ee 
where the function $\Omega_k(\tau)$ admits an asymptotic  expansion in terms of the derivatives of $a(\tau)$
\be \label{adiabaticexpansionN}\Omega_k = \omega+ \omega_k^{(1)}+ \omega_k^{(2)} + \omega_k^{(3)}+ \omega_k^{(4)} + \cdots   \ . \ee
The coefficients of the expansion $\omega_k^{(n)}$ are obtained by systematic iteration from the mode equation, and depend on derivatives of $a$ up to and including order $n$. The expansion above dictates the ultraviolet behavior that the fields modes must obey in order to define an admissible  quantum state. Note that \eqref{conformalmodes} and \eqref{adiabaticexpansionN} should be satisfied at all orders to be equivalent to the Hadamard condition.
This expansion is in general  asymptotic, and therefore cannot define a unique vacuum state, but rather a family of acceptable states. 
Furthermore, the two-point function inherits from \eqref{conformalmodes} and \eqref{adiabaticexpansionN}  an adiabatic expansion %{\color{blue}\sout{ which is equivalent to the DeWitt-Schwinger expansion  when it is restricted to  FLRW spacetimes } 
which produces the same renormalized stress-energy tensor as the DeWitt-Schwinger expansion when it is restricted to  FLRW spacetimes in four spacetime dimensions \cite{beltran-nadal1,rio-navarro,Silvia-Winstanley}.  In Appendix \ref{ap:adiabatic} we give more detail about the adiabatic method for scalars.  \\

For the $CPT$-invariant vacua, parametrized by the real and time-independent function $\theta_k$, this condition should be reexpressed in terms of the hyperbolic initial phase.  It is important to note that the adiabatic modes $\varphi_k^{(N)}$ of order $N$ satisfy the equation of motion at order $N$. Therefore, adiabaticity is preserved in time and it is enough to study the large momentum behavior of the modes $\varphi_k$ at a given instant of time $\tau_0$ to study its ultraviolet behavior.\\

In the context of our analysis, it is natural to evaluate the complete adiabatic expansion of the modes \eqref{conformalmodes} at $\tau=0$.  We get 
\be \label{aexpansion}\varphi_k (0) \sim \frac{1}{\sqrt{k}}+\frac{\gamma^2}{8\,k^{9/2}}+\frac{41 \gamma^4}{128\, k^{17/2}}+\cdots\, .\ee
From this, we infer an asymptotic expansion for  $\theta_k$ 
\be \label{aexpansiontheta}\theta_k \sim \frac{\gamma^2}{8 k^4}+\frac{5 \gamma^4}{16 k^8}+\frac{61\gamma^6}{24 k^{12}}+\cdots \, . \ee
The set of vacuum states that fit the above large $k$ expansion can be generically referred to as {\it adiabatic ($CPT$-invariant) vacua}. %It is important to note that, although we have imposed the adiabatic condition at $\tau=0$, the adiabaticity of the vacuum state will remain unaltered in time {\color{red}[dir aquesta frase d'altra forma, ja que en la imatge de Heisenberg el buit es time-independent]}. 
 In Appendix \ref{ap:expansions} we show that only if $\theta_k$ behaves as in \eqref{aexpansiontheta}, the high-energy behavior of the field modes is compatible with the adiabatic condition after time evolution (i.e., for $\tau>0$).

\section{ States of Low Energy for scalars}\label{sec::SLE::Scalars}
%{\color{red}[ser consistent en la notació $\to$ discutir en el proxim meeting]}

Let us briefly summarize the method for constructing states of low energy (SLE). We follow the prescription described in \cite{Olbermann07, Banerjee20}. The main idea of this construction is to fix the free parameters of the problem (e.g., $\theta_k$ in our $CPT$-invariant model) by requiring that the vacuum expectation value of the smeared energy density over a temporal window should be minimal. A special and very important virtue of this construction is that it guarantees that the resulting states are  Hadamard for smooth regions of the spacetime. We first review the general method for arbitrary $a(\tau)$ emphasizing the aspects that will be relevant for our analysis and then we particularize it for the $CPT$-invariant model under consideration. We further generalize the standard result by including an arbitrary coupling to the scalar curvature $\xi$. The starting point is to fix a fiducial set of normalized modes $\phi_k(\tau)$. They are related to the Weyl modes of the previous section by $\varphi_k= a\, \phi_{k}$. One can parametrize a general set of modes $T_k(\tau)$ in the form
\be \label{TS}T_k(\tau) = \lambda_k \phi_k(\tau) + \mu_k \phi_k^*(\tau) \ , \ee
where $\lambda_k$ and $\mu_k$ are complex numbers that must obey $|\lambda_k|^2 -|\mu_k|^2 =1$. The goal now is to find for which values of $\lambda_k$ and $\mu_k$ the smeared energy density is minimal. \\
 
The smeared energy density can be defined as follows. The vacuum expectation value of the Hamiltonian $H(\tau)$ for a  scalar field $\phi$, associated to a foliation of a spacelike Cauchy surface $\Sigma_{\tau}$ and a temporal vector characterizing an observer $u^b$ is given by
\be
 H(\tau) =\int_{\Sigma_{\tau}}\textrm{d}^{3}x \sqrt{|h|}\, \langle T_{ab}\rangle\, n^{a} u^{b}
\ . \ee
$\sqrt{|h|}$ is the determinant of the induced Riemann metric in $\Sigma_{\tau}$,  $\langle T_{ab}\rangle $ is the vacuum expectation value of the stress-energy tensor, $n^{a}$ is the unit normal vector to the foliation.  We should take as $\langle T_{ab} \rangle$  the renormalized values. However, we are interested in the problem of minimizing the energy density, and since the subtraction terms in the renormalization are independent of the vacuum state we can ignore them. We will only consider the contributions of the modes, as in \eqref{modesrhok}.  %In a FLRW spacetime usually 
Following \cite{Olbermann07}, we choose the privileged isotropic observer, $u^{a} = n^{a} =(a^{-1},0,0,0)$ in conformal time coordinates, and the above expression reduces to
\be\label{Hamiltonian}
H(\tau) =\int_{\Sigma_{\tau}}\textrm{d}^{3}x \sqrt{|h|}\, \langle T_{ab} \rangle\, u^{a} u^{b}= \int_{\Sigma_{\tau}}\textrm{d}^{3}x \sqrt{|h|} \,\langle \rho(\tau)\rangle  = \int_{\Sigma_{\tau}}\textrm{d}^{3}x \int \frac{\textrm{d}^3k}{(2\pi)^{3/2}} \sqrt{|h|}\, \rho_k(\tau) \, ,
\ee
where %$\textrm{d}^3k'=\frac{\textrm{d}^3k}{\sqrt{(2\pi)^3}}$ and 
 $\rho_k(\tau)$ is the formal energy density of a given mode $T_k$. In  conformal coordinates $\rho_k(\tau)$ is given by %\cite{Ferreiro:2022hik}
\be \label{modesrhok}
\rho_k(\tau)=\frac{1}{4a^{2}}\left(|T'_k|^2+\omega^2 |T_k|^2+6 \xi\big(\frac{a^{\prime 2}}{a^2}\left|T_k\right|^2+\frac{a^{\prime}}{a}\left(T_k  T_k^{\prime*}+ T_k^* T_k^{\prime}\right)\big)\right)\, ,
\ee
(remember $\omega^2=k^2 + m^2 a^2$). It is interesting to stress that, for a radiation-dominated universe, $T_k \sim a^{-1}$ and  therefore the behavior of $\rho_k$ near the big bang is  $\rho_k \sim a^{-6}$. This will be important in the following section.\\

The smeared Hamiltonian is defined then as 
\be
 H[f]  :=\int \textrm{d}\tau \sqrt{a^{2}}\, f^2(\tau) H(\tau)  =\int \frac{\textrm{d}^3k}{(2\pi)^{3/2}}\int_{\Sigma_{\tau}}\textrm{d}^{3}x\int \textrm{d}\tau  \sqrt{|g|} \,f^2(\tau)  \rho_k(\tau) \, , 
\ee
 where we have used \eqref{Hamiltonian}, and where the smearing has been done with a positive definite window function $f^2$, along the curve of an isotropic observer. One can see that the temporal dependence term to be minimized is indeed the smeared energy density for each mode $k$ with the appropriate factor $\sqrt{|g|}$ coming from the four-dimensional volume element. Therefore, from now on we will work directly with the smeared energy density %From now on we will be working directly with  
\bea \label{eq:smearedED}
\mathcal{E}_k[f] &:=& \int \text{d} \tau\, \sqrt{|g|}\, f^2 \, \rho_k \, .
\eea
As stated above, the SLE prescription is based on choosing the (Hadamard) state that minimizes the energy density over a temporal window function \eqref{eq:smearedED}. It is important to stress the appearance of $\sqrt{|g|}$ which %for any FLRW spacetime
in our case reduces to $\sqrt{|g|}= a^{4}$ . We note that this type of factors also appear in the analysis of Ref. \cite{ADLS2021} to smooth the big bang singularity via quantized fields. %This term smooths the big bang singularity as argued in \cite{ADLS2021}.  
In Sec. \ref{Subsec::BigBang::Scalars} we will apply this prescription to build a Hadamard state around $\tau=0$ in a radiation-dominated spacetime by taking advantage of this volume element.\\

In order to minimize $\mathcal{E}_k$, it is very convenient to express it explicitly in terms of the free parameters $\mu_k$ and $\lambda_k$, namely
\be
\mathcal{E}_k=(2 \mu_k^2+1)c_{k,1}+2 \mu_k \textrm{Re}(\lambda_k c_{k,2})\, ,
\ee
where we have defined 
\bea \label{eq:c1_conformal_Weyl}
c_{1}&\equiv&c_{k,1}=\frac{1}{4} \int  \text{d} \tau\, \sqrt{|g|}\,\frac{f^2}{a^2}\left(|\phi'_k|^2+\omega^2|\phi_k|^2+6 \xi\big(\frac{a^{\prime 2}}{a^2}\left|\phi_k\right|^2+\frac{a^{\prime}}{a}\left(\phi_k  \phi_k^{\prime*}+\phi_k^* \phi_k^{\prime}\right)\big)\right) \, ,\\
c_2&\equiv&c_{k,2}=\frac{1}{4} \int \text{d} \tau \sqrt{|g|}\,  \frac{f^2}{a^2}\left(\phi'_k{}^2+\omega^2 \phi_k^2+6 \xi\big(\frac{a^{\prime 2}}{a^2}\phi_k^2+2\frac{a^{\prime}}{a}\phi_k  \phi_k^{\prime}\big)\right)\, . \label{eq:c2_conformal_Weyl}
\eea
 In the above formulas it is assumed that $c_1$ is a positive quantity, as it is trivially satisfied for $\xi=0$. It can be showed that if we take $\mu_k$ to be real and positive, the minimization problem over the parameters $\lambda_k$ and $\mu_k$ determines a unique solution, namely
\be\label{mu-lambda} \mu_k= \sqrt{\frac{c_{1}}{2\sqrt{c_{1}^2 -|c_2|^2}} -\frac{1}{2}} \ ; \qquad \lambda_k=-e^{-i \text{Arg} \ c_2}  \sqrt{\frac{c_{1}}{2\sqrt{c_{1}^2 -|c_2|^2}} +\frac{1}{2}} \ . \ee
Notice that the minimization problem holds whenever $\frac{|c_2|}{c_{1}}<1$ is satisfied. This  is usually the case if $\phi_k$ do not contain singularities in the support of $f^2$ \cite{Olbermann07, Banerjee20}. This issue will be relevant around the big bang singularity, as we will see in the following sections.\\ %We remark that the resulting state of low energy is independent of the fiducial solution $\phi_k$ and it is a Hadamard state (i.e., a state of infinite adabatic order).\\

\subsection{$CPT$-invariant States of Low Energy}
Let us now extend the method above for the $CPT$-invariant states in a radiation-dominated universe. If the minimizing problem is restricted to the set of $CPT$-invariant states, we can parametrize the state of low energy using $\theta_k$. For this we choose as a fiducial solution
\be \label{eq:basis-CPT-scalars}
\phi_k(\tau)=\frac{1}{a}\varphi^{CPT}_k(\tau,\theta_k=0)\, , \ee with $\varphi_k^{CPT}$ given in \eqref{eq:sol-scalars-cpt} and choosing $\theta_k=0$. If we impose CPT to our general solution \eqref{TS}, using the CPT conditions at $\tau=0$, [i.e., $\varphi_k(0) =  \varphi_k^*(0)$ and $\varphi'_k(0) =  -\varphi'{}_k^*(0)$] we arrive to 
\be \label{S-proposal}
\lambda_k=\cosh(\theta_k)\, , \qquad \mu_k=\sinh(\theta_k)\, .
\ee
That is, $\lambda_k$ and $\mu_k$ are real functions. Now we proceed to minimize $\mathcal{E}_k$ with the $\theta_k$ parametrization
\be
\mathcal{E}_k= (2 \mu^2_k+1)c_{1}+2 \mu_k \textrm{Re}(\lambda_k c_2) \quad \rightarrow \quad \mathcal{E}_k=c_{1} \cosh(2\theta_k)+\sinh(2 \theta_k)\textrm{Re}(c_2)\, ;
\ee
Taking $\partial_{\theta_k} W=0$ we end up with
\be\label{eq:final-tanh}
\tanh(2\theta_k)=-\frac{\textrm{Re}(c_2)}{c_{1}}\, .
\ee
So the state given by
\be \label{eq:T-minimal-scalars}
T_{k}= \cosh{(\theta_k)} \phi_{k} +  \sinh{(\theta_k)} \phi^{*}_{k}\, ,
\ee
with $\phi_k$ given in \eqref{eq:basis-CPT-scalars} and
\be\label{CPT-SLE}
\theta_{k}= \frac{1}{2} \text{arctanh}\left(- \frac{\text{Re}(c_2)}{c_{1}}\right)
\ee
corresponds to the $CPT$-invariant state of low energy. Note again that, although a particular fiducial solution has been chosen in the minimization process, the final result \eqref{eq:T-minimal-scalars} is independent of this basis.  However it depends, in general, on the choice of $f^2$.

\subsubsection{An example: $CPT$-invariant vacuum of low energy at late times}\label{Subsec::LateTimes::Scalars}

Let us see how to obtain an example of an adiabatic (Hadamard) vacuum which is also $CPT$-invariant using the prescription above. For each choice of the initial vacuum $|0 \rangle$ we have a different prediction for the particle creation spectrum at late times. Particles are defined at $\tau \to +\infty$  as excitations of the {\it out-}vacuum $|0_+\rangle$. The particle production rate can be obtained by the frequency-mixing approach \cite{Parker66,Parker68,Parker69,Parker71}. It has been reviewed in \cite{parker-toms, birrell-davies, fabbri-navarro, ford21} and used extensively in the literature for decades (see, for instance, \cite{PF,Mamaev,Audretsch78,Ford, Chung, Anderson-Mottola-Sanders, Good, Ema, BP-F-N-P, Xue}). In our case, we find  the following expression  for the average density number of created particles in the mode $\vec k$ as a function of the initial state characterized by $\theta_k$,

\be \label{particle-num-theta}
    n_k \equiv|\beta_k|^2=-\frac{1}{2}+\frac{e^{-\pi \kappa}\cosh{(2 \pi \kappa)}}{4 \pi }\Bigg( e^{-2\theta_k}\,\kappa^{\frac{1}{2}}\,|\Gamma(\tfrac{1}{4}+i\kappa)|^2+e^{2\theta_k} \,\kappa^{-\frac{1}{2}}\,|\Gamma(\tfrac{3}{4}+i\kappa)|^2\Bigg)\, ,
    \ee
with $\kappa$ given in Eq. \eqref{eq:kappa:k}. We remark that the above general expression can be reexpressed, after some manipulations, as
    \be
    |\beta_k|^2=\frac{1}{2}\Big(\cosh(2 \eta_k) \cosh( \Lambda_k)-1\Big)
    \ , \ee
    where we have defined $\cosh(\Lambda_k)=\sqrt{1+e^{-4 \pi \kappa}}$, and  
 \be 2 \eta_k =2 (\theta_k^{late}-\theta_k)\, ,\ee  with
    \be \label{thetakT}
    \theta_k^{late}=\frac{1}{4}\ln\Big(\frac{\kappa\cosh{(2 \pi \kappa)}}{2 \pi^2} |\Gamma(\tfrac{1}{4}+i\kappa)|^4\Big)\,.
  \ee
The state producing the minimal amount of particles at late time is therefore given by
\be \theta_k= \theta_k^{late} \ . \ee
This state can be seen as a $CPT$-invariant low energy state associated to a smearing function with support at $|\tau| \sim \infty$ since at late times (here we include the conventional renormalization subtractions),
\be
\mathcal{E}_k\left[f\right] \propto \frac{1}{2} \int \text{d} \tau\, f^2\,  \omega\,|\beta_k|^2 \, .
\ee
It is important to remark that, in this case, this state  minimizes the smeared energy density at late times independently of the choice of smearing function $f^2$. % According to \cite{Olbermann07}, low energy states are  Hadamard states. As expected, this state is a Hadamard state. 
One can check that this state is Hadamard by evaluating the asymptotic large $k$ expansion of $\theta_k^{late}$ 
    \be \label{thetakT-late}
    \theta_k^{late}=\frac{1}{4}\ln\Big(\frac{\kappa\cosh{(2 \pi \kappa)}}{2 \pi^2} |\Gamma(\tfrac{1}{4}+i\kappa)|^4\Big) \sim \frac{\gamma^2}{8 k^4}+\frac{5 \gamma^4}{16 k^8}+\frac{61\gamma^6}{24 k^{12}}+\cdots \,  \ee
and comparing with the adiabatic expansion (\ref{aexpansiontheta}).  %It agrees with the adiabatic expansion (\ref{aexpansiontheta}) at any order.  
As expected, they agree at all orders. This Hadamard and $CPT$-invariant state is  equivalent to the one proposed in \cite{LetterCPT, Large}. We note that, if we do not impose CPT invariance, the states of low energy at early and late times are just $|0_\pm \rangle$, which have zero energy [see Eqs. \eqref{eq:adiabatic-late-scalars} and \eqref{eq:adiabatic-early-scalars}]. Finally, we also point out that this identification of the low energy state at late times is somewhat similar to the characterization of the Bunch-Davies vacuum as the low energy state in de Sitter  when the smearing function has support at very early times. This is also independent of the particular smearing function  \cite{Olmedo, Degner2010}.\\

In the following section we will study states of low energy in a CPT-symmetric radiation-dominated universe but with $f^2$ supported around the big bang ($\tau=0$). As we will see, in this case we find that the result strongly depends on the choice of $f^2$, and also on $\xi$. For $\xi \neq \frac{1}{6}$ the issue is more subtle, and an extra condition has to be imposed on $f^2$. For $\xi=\frac{1}{6}$ this constraint is alleviated.

\section{$CPT$-invariant States of Low Energy at $\tau=0$ for scalars }\label{Subsec::BigBang::Scalars}

In the last subsection, we have obtained a $CPT$-invariant state by minimizing the smeared energy density at  $|\tau| \to \infty$. However, this may seem an unnatural vacuum state since we are imposing an initial condition $\theta_k=\theta_k^{late}$ which is determined by the late time behavior of the Universe. %behavior of the modes. 
Therefore, a simple question arises: can we obtain a Hadamard state by minimizing the smeared energy density supported around the big bang, $\tau = 0$ ? We will study the above question using a Gaussian smearing function
\be\label{Gaussian}
f_g^2(\tau)= \frac{1}{\sqrt{\pi}\epsilon}e^{-\frac{\tau^2}{\epsilon^2}}\, .\ee
%{\color{green} as it is also done in \cite{Degner2010}.}
%This function belongs to the Schwartz space. The behavior of the state with respect to the big bang singularity has also been studied with other functions with compact support around $\tau=0$ rendering similar results.} %{\color{blue} Potser llevar açò i comentar que és una funció de l'espai de Schwartz i prou, no cal que siga de suport compacte per al que estem analitzant nosaltres, però està bé saber que el resultat és equivalent per a suports compactes. The behavior of the state with respect to the big bang singularity has also been studied with other functions with compact support around $\tau=0$ rendering the same results. Therefore, \eqref{Gaussian} will be enough for our analysis.}\\

In this section we will work with minimally coupled scalars $\xi=0$ for simplicity. The results can be generalized to other couplings except for the very special conformal coupling $\xi=1/6$. We detail this last for completeness in Appendix \ref{App::SLE::Conf}.

\subsection{Massless case}

Let us first analyze the massless case, where we can find analytic solutions. %We will work with the Weyl modes $\varphi_k(\tau)=a(\tau)\phi_k(\tau)$.  
We take the fiducial solution given in \eqref{eq:basis-CPT-scalars}, that in this case corresponds to the conformal solution $\phi_k= \frac{e^{-i k \tau}}{a\sqrt{k}}$. We proceed to obtain the $CPT$-invariant state of low energy with the Gaussian function centered at $\tau=0$, i.e., $f^2= f_g^2$. 
We first evaluate $c_{1}$ and $c_2$
\bea \label{eq:c1_c2_conformal_Weyl_Massless1}
c_{1} && =\frac{1}{4}  \int \text{d} \tau\, f_g^2 \left( \frac{1}{2 k \tau^2}+ k \right) = \frac{k}{4}-\frac{1}{4 k \epsilon ^2}\, , \\
c_2 && =\frac{1}{4}\int \text{d} \tau\, f_g^2 \, e^{-2i k \tau} \left(  \frac{1}{2 k \tau ^2}+\frac{i}{\tau }  \right)= -\frac{e^{-k^2 \epsilon^2}}{4 k \epsilon^2} \ . \label{eq:c1_c2_conformal_Weyl_Massless2}\eea
We note that, in evaluating them, we made use of the distributional character of the integrand (see, for example, \cite{ADLS2021}). However, although this quantities give finite results, $c_{1}$ changes sign depending on the value of $k$, and therefore the quotient $\frac{|c_2|}{c_{1}}$
is not necessarily smaller than 1 for all $k$, thus the minimization prescription cannot be applied around $\tau=0$ whenever we have a divergence in the above integrals. To bypass this problem and to be able to minimize the smeared energy density for all $k$ we require the  condition %a smearing function that behaves as %$f^2= a^{2}\, f_g^2$ 
\be \lim_{\tau\to 0} \frac{f^2(\tau)}{\tau^2}<\infty\, .\ee
 For example, using $f^2=a^2\, f_g^2$ one obtains the following state of low energy for all $k$ centered at $\tau=0$
\be \theta_{k}^{m=0}= -\frac{1}{2} \coth ^{-1}\left(e^{(\epsilon k) ^2}\frac{1+(\epsilon k) ^2}{1+2 (\epsilon k)
   ^2}\right) \, . \ee
The resulting state is Hadamard because the large momentum expansion of $\theta_k$ decays faster than any power of $k^{-n}$ [see Eq. \eqref{aexpansiontheta} for $\gamma=0$].  The divergent terms $\tau^{-2}$ in the integrands of Eqs. \eqref{eq:c1_c2_conformal_Weyl_Massless1} and \eqref{eq:c1_c2_conformal_Weyl_Massless2} disappear for a conformally coupled field (see Appendix \ref{App::SLE::Conf}) and the $\sqrt{|g|}=a^{4}$ factor from the volume element is enough to render both integrals finite. However, because we are minimizing the state around the big bang, the resulting state depends on the test function $f^2$.

\subsection{Massive case}

 Let us now study the massive case. In this context we cannot obtain an analytic expression for the state of low energy centered at $\tau=0$. However, we can obtain an approximated state  given by the expansion of the modes $\phi_k$ around $\tau \sim 0$, and study its large-$k$ behavior. %by expanding %the energy density part for $\phi_k(\tau)$ in 
 We have explicitly checked that this approximated solution for $\theta_k$ follows the adiabatic condition \eqref{aexpansiontheta} up to a given order, that increases as we increase the order of the expansions  in powers of $\tau$. \\

This analysis can be done as follows. We start from the definition of $c_1$ and $c_2$ given in Eqs. \eqref{eq:c1_conformal_Weyl} and \eqref{eq:c2_conformal_Weyl} with the fiducial modes $\phi_k$ proposed in Eqs. \eqref{eq:basis-CPT-scalars}, and using $f^2= a^2\,f_g^2$ as the smearing function to ensure a well posed minimization problem. %An approximated solution for the Hadamard state will be given if 
We then expand the modes in powers of $\tau$ around $\tau=0$ (where the center of the temporal window is located), and find %the support of the temporal window, which in this case is centered at $\tau=0$. %Therefore for $c_{1}$ and $c_2$ we have
\bea
c_{1} &&=\frac{1}{4}  \int \text{d} \tau\, a^{2}\,f_g^2 \left( \frac{1}{k \tau ^2}+2k +\frac{3 \gamma ^2 \tau ^2}{2
   k} -\frac{1}{9} \gamma ^2 k \tau ^4 \, \cdots  \right)\,  ,\label{c1exp}\\
c_2&&=\frac{1}{4}  \int \text{d} \tau\, a^2\, f_g^2   \left(\frac{1}{k \tau ^2}+2k-\frac{8}{3} i k^2 \tau -2 k^3 \tau ^2+\frac{3 \gamma ^2 \tau^2}{2 k} %-\frac{12}{5} i \gamma ^2 \tau ^3+\frac{16}{15} i k^4 \tau ^3
\, \cdots  \right). \label{c2exp}
\eea 
We remark that we have chosen $f^2= a^2 f_g^2$ for the smearing function so that the minimization problem is well posed for all $k$. \\%\\We have computed the above integrals up to the order $O\left(\tau^{14} \right)$.\\ 

Finally, using Eq. \eqref{CPT-SLE}, we take the expansions above and use them to obtain an approximated (and very involved) expression for the initial phase $\theta_k$. As a byproduct, we can use this result to check that the state obeys the Hadamard condition. Namely, if we expand our result for large $k$ we recover order-by-order the expected asymptotic behavior \eqref{aexpansiontheta}. As we include more terms of the $\tau$ expansion, more terms in the large $k$ expansion are recovered. %in \eqref{c1exp} and \eqref{c2exp}. 
We have explicitly proved that for $\mathcal{O}\left(\tau^{14}\right)$ in \eqref{c1exp} and \eqref{c2exp} we obtain the expected large momentum expansion up to $ \mathcal{O}\left( k^{-16} \right)$. 
\be \label{thetakbigbang}
    \theta_k \sim \frac{\gamma^2}{8 k^4}+\frac{5 \gamma^4}{16 k^8}+\frac{61\gamma^6}{24 k^{12}}+ \mathcal{O}\left( k^{-16} \right) \, . \ee
These computations involves very long analytical expressions and have required the intensive use of the \emph{Mathematica} software.\\

In summary,  with the prescription of states of low energy we can obtain a $CPT$-invariant Hadamard state which minimizes the smeared energy density around the big bang for a smearing function $f^2$ %that satisfies %of the kind
%{\color{red}\be\lim_{\tau\to 0} \frac{f^2(\tau)}{\tau^2}< \infty\, ,\ee
such as $f^2=a^{2}\,f_g^2$. We note that the asymptotic expansion \eqref{thetakbigbang} is not sensitive to the particular smearing function $f^{2}$ that we are using: the particular choice of  $f^{2}$ only matters in the infrared regime. %This ambiguity  included in the choice of smearing function could be naturally interpreted, at least heuristically, as encoding quantum gravity effects.
We also note that the result above \eqref{thetakbigbang} is {\it independent of the scalar coupling $\xi$}. We have explicitly checked that the large $k$ expansion of the hyperbolic angle $\theta_k$ obtained via the SLE prescription does not depend on $\xi$.

\section{$CPT$-invariant states for fermions in a radiation-dominated spacetime}

\label{sec:spinor-CPT}

Let us consider now a spin one-half field $\Psi$ propagating in the same background metric $\text{d}s^2= a^2(\tau)(\text{d}\tau^2 - \text{d}\vec x^2)$. The field equation 
$(i \underline \gamma^\mu\nabla_\mu-m)\Psi=0
$, 
where $\underline \gamma^\mu=\frac{1}{a}\gamma^\mu$ and $\gamma^\mu$ are the flat spacetime Dirac matrices, become\footnote{We use the conventions of \cite{parker-toms, birrell-davies}.}
\be
\Big(\gamma^0\partial_\tau + \vec \gamma \cdot \vec \nabla +\frac{3 a'}{2 a} \gamma^0+im a\Big)\Psi=0
\ . \ee
It is also convenient to  perform a Weyl transformation for the spinor field of the form $\psi= a^{3/2}\Psi$. The  mode expansion for the quantized $\psi$ field is given by ($D_{\vec{k} h}=B_{\vec{k} h}$ for Majorana spinors)
\be\label{psiexpansion}
\psi(x)=\int \text{d}^{3}k \sum_{h}\left[B_{\vec{k} h} u_{\vec{k} h}(x)+D_{\vec{k} h}^{\dagger} v_{\vec{k} h}(x)\right]\, ,
\ee
where the subindex $h$  refers to  the  helicity, and where the $u$-modes in the Dirac representation 
\be
\gamma^{0}=\left(\begin{array}{cc}
I & 0 \\
0 & -I
\end{array}\right), \quad \vec{\gamma}=\left(\begin{array}{cc}
0 & \vec{\sigma} \\
-\vec{\sigma} & 0
\end{array}\right) \, ,
\ee
can be written  as (we have reexpressed the results in \cite{LNT1,LNT2} in terms of the conformal time, up to the above $a^{-3/2}$ Weyl rescaling factor)

\be \label{udefinition}
u_{\vec{k} h}(x)=\frac{e^{i \vec{k} \cdot \vec{x}}}{\sqrt{(2 \pi)^{3} }}\left(\begin{array}{c}
h_{k}^{I}(\tau) \xi_{h}(\vec{k}) \\
h_{k}^{I I}(\tau) \frac{\vec{\sigma} \cdot \vec{k}}{k} \xi_{h}(\vec{k})
\end{array}\right)\, , \ee
and where   $\xi_{h}$ is a constant and normalized two-component spinor $\xi_{h}^{\dagger} \xi_{h^{\prime}}=\delta_{h^{\prime} h}$ that represent the helicity eigenstates. $v_{\vec{k} h}(x)$ is obtained by the charge conjugation operation. The Dirac equation for the functions $h^I_k (\tau)$ and $h_k^{II}(\tau) $ is transformed into [we define again $m a=\gamma \tau$]

\bea \label{modeh10}
h^{\prime I}_k+i k \,h^{II}_k+i\, \gamma \tau \,h^{I}_k=0\, ,\\ \label{modeh20} 
h^{\prime II}_k+i k\, h^{I}_k-i\, \gamma \tau\, h^{II}_k=0\, ,
\eea
together with the normalization condition
\be\label{norm-fermions}
|h_k^{I}|^2+|h_k^{II}|^2=1\, .
\ee
 As in the scalar case, the general  solution can be given in terms of parabolic cylindrical functions $D_{\nu}(z)$ as
  \bea \label{h1rad}
 h^{I}_k&=&C_{k,1}\,  s^{I}_k(\tau)+  C_{k,2}\, s_k^{II*}(-\tau)\,,\\
 h^{II}_k&=&C_{k,1} \, s_k^{II}(\tau)+C_{k,2}\, s^{I*}_k(-\tau),\label{h2rad}
 \eea
where 
\be
s_k^{I}=\frac{1}{\sqrt{2}}D_{-2 i \kappa}(e^{\frac{i \pi}{4}}\sqrt{2 \gamma}\, \tau)\, , \qquad s_k^{II}= e^{\frac{i \pi}{4}}\sqrt{\kappa }\,D_{-1-2 i \kappa}(e^{\frac{i \pi}{4}}\sqrt{2 \gamma}\, \tau)\, ,
\ee
and where again $\kappa=\frac{k^2}{4 \gamma}$. The complex functions $C_{k,1}$ and $C_{k,2}$ defining the vacuum state are constrained by the normalization condition \eqref{norm-fermions}.\footnote{In terms of $C_{k,1}$ and $C_{k,2}$ the normalization condition reads $$ \frac{e^{\pi\kappa }}{2}(|C_{k,1}|^2+|C_{k2}|^2)+\frac{2\sqrt{\kappa }\,\sinh{(2 \pi \kappa)}}{\sqrt{\pi}}\textrm{Re}[e^{-i\frac{\pi}{4}}C_{k,1}C_{k,2}^*\Gamma(-2i\kappa)]=1\, .$$}\\ 

It is important to note that
 the  equations (\ref{modeh10}) and (\ref{modeh20}) for the time-dependent part of the field modes   $h^{I}_k$ and $h^{II}_k$ remain unchanged under the transformation $\tau \to -\tau$ and, simultaneously, $h_k^I\to h_k^{II*}$ and $h_k^{II}\to h_k^{I*}$. %This is indeed the form of the CPT transformation given in \cite{Large} on the above characterization of the field modes. 
 Therefore the vacuum will be $CPT$-invariant if the chosen modes  verify the relation \bea  h_k^I(\tau)= h_k^{II*}(-\tau) \, . \eea \\
% h_k^{II}(\tau) = h_k^{I*} (-\tau) \ . \eea
 In terms of the functions $C_{k,1}$ and $C_{k,2}$ a $CPT$-invariant vacuum is characterized by the restriction $C_{k,1}=C_{k,2}^{*}$. As for the scalar field, we can easily constraint the $CPT$-invariant initial conditions at $\tau=0$ as
\bea \label{cpt0}h^I_k(0)= h^{II*}_k(0) 
  \ . \eea
Furthermore, we have also the normalization condition (\ref{norm-fermions}) at $\tau=0$
%\be \label{Normalization0}|h^{I}_k(0)|^2+|h^{II}_k (0)|^2=1 \ , \ee
implying  \be \label{Normalization1} |h_k^{I}(0)|=|h_k^{II}(0)|=\frac{1}{\sqrt{2}} \ . \ee
Therefore, the general ($\tau=0$) solution to the conditions (\ref{cpt0}) and (\ref{Normalization1}) can be written as  
\bea \label{eq:initial-Dirac}
h^{I}_k(0)=\frac{e^{+i\Theta_k}}{\sqrt{2}} \, , \qquad  
h^{II}_k(0)=\frac{e^{-i\Theta_k}}{\sqrt{2}}
\ , \eea
where $\Theta_k$ is an arbitrary trigonometric angle. 
In terms of $\Theta_k$, the constants $C_{k,1}$ and $C_{k,2}$ read
\be \label{eq:C1-fermions}
C_{k,1}=2^{i \kappa}\sqrt{\pi}  e^{\pi \kappa}\Bigg(\frac{e^{i \Theta_k}}{\Gamma(\frac{1}{2}-i \kappa)}+\frac{\kappa^{\frac{1}{2}}e^{i\frac{3\pi}{4}}\,e^{-i \Theta_k}}{\Gamma(1-i\kappa)}\Bigg)\, ,
\ee
 $C_{k,2}=C_{k,1}^*$. Therefore, the $CPT$-invariant solution reads
  \bea \label{h1rad-cpt}
 h^{I;CPT}_k&=&C_{k,1}\,  s^{I}_k(\tau)+  C_{k,1}^*\, s_k^{II*}(-\tau)\,,\\
 h^{II;CPT}_k&=&C_{k,1} \, s_k^{II}(\tau)+C_{k,1}^*\, s^{I*}_k(-\tau),\label{h2rad-cpt}
 \eea
and with $C_{k,1}$ given above.\\

%\subsubsection{Late-times vacuum}

As for the scalar field case, it is also possible to find solutions that are not $CPT$-invariant. In particular, at late times there is a preferred solution for the field modes given by the leading order adiabatic expansion (positive-frequency solution)%. This solution is given by the late-times asymptotic condition
\bea
h^{I\,(+)}_k(\tau)&\sim& \sqrt{\frac{\omega+m a}{2 \omega}} e^{-i \int_\tau \omega(u)du}\sim e^{-i\left(\frac{\gamma}{2}\tau^2+\kappa \ln(2 \gamma \tau^2)\right)}\, ,\\
h^{II\,(+)}_k(\tau)&\sim&\sqrt{\frac{\omega-m a}{2 \omega}} e^{-i \int_\tau \omega(u)du}\sim \frac{\sqrt{\kappa}}{\sqrt{\gamma}\,\tau}e^{-i\left(\frac{\gamma}{2}\tau^2+\kappa \ln(2 \gamma \tau^2)\right)}\, ,
\eea
that leads to
\be
C_{k,1}=\sqrt{2} e^{-\frac{\pi \kappa}{2}}\, , \qquad C_{k,2}=0\, .
\ee
This solution is not $CPT$-invariant since $C_{k,1}\neq C_{k,2}^{*}$. For completeness, we also give the form of the preferred solution at  early-times ($\tau\to -\infty$). Again, the solution is fixed by imposing the late-times negative-frequency behviour
\bea
h^{I\,(-)}_k(\tau)&\sim& \sqrt{\frac{\omega+m a}{2 \omega}} e^{-i \int_\tau \omega(u)du}\sim -\frac{\sqrt{\kappa}}{\sqrt{\gamma}\,\tau}e^{+i\left(\frac{\gamma}{2}\tau^2+\kappa \ln(2 \gamma (-\tau)^2)\right)}\, ,\\%e^{-i\left(\frac{\gamma}{2}\tau^2+\kappa \ln(2 \gamma \tau^2)\right)}\\
h^{II\,(-)}_k(\tau)&\sim&\sqrt{\frac{\omega-m a}{2 \omega}} e^{-i \int_\tau \omega(u)du}\sim  e^{+i\left(\frac{\gamma}{2}\tau^2+\kappa \ln(2 \gamma (-\tau)^2)\right)}\, .
\eea
From this condition we get $C_{k,1}=0$ and $C_{k,2}=\sqrt{2} e^{-\frac{\pi \kappa}{2}}$. 

\subsection{Ultraviolet regularity of the $CPT$-invariant vacuum states}

As happens with the scalar field case, it becomes fundamental to study the ultraviolet regularity of the $CPT$-invariant vacuum states. For cosmological backgrounds and for spin-$\frac{1}{2}$ fields it means that for large $k$, the behavior of the modes $h^{I}_k(\tau)$ and $h_k^{II}(\tau)$ should be dictated by their adiabatic expansion. The analysis of the adiabatic expansion for spinors is more involved than for scalars. It does not fit the conventional WKB-type template, as happens for scalar fields. It is given, assuming  the definitions (\ref{udefinition}) for the modes, by \cite{LNT1,LNT2, DNT,BFNV}
\bea \label{eq:adiabatic:fermions}
h^{I}_k(\tau) &\sim& \sqrt{\frac{\omega + m a }{2 \omega}}\,\big(1 + F_k^{(1)} + F_k^{(2)} + \cdots\big)\, e^{-i\int^\tau \Omega_k(\tau') d\tau'} \ ,   \\
h^{II}_k (\tau) &\sim&\sqrt{\frac{\omega -ma }{2 \omega}}\,\big(1 + G_k^{(1)} + G_k^{(2)} + \cdots\big)\,e^{-i\int^\tau \Omega_k(\tau')  d\tau'}  \ , \nonumber
\eea
where again $\omega^2= k^2 + m^2a^2$, and $\Omega_k(\tau) = \omega + \omega_k^{(1)} +\omega_k^{(2)} +\cdots$. The recursive algorithm is displayed  in \cite{LNT1,LNT2, DNT,BFNV}. Note that \eqref{eq:adiabatic:fermions} provides the adiabatic condition for spin-$\frac{1}{2}$ fields: for large $k$ the behavior of the modes $h^{I,II}_k$ should follow this expansion at all orders. This is the analogous adiabatic condition for scalar fields defined in Eqs.  \eqref{conformalmodes} and \eqref{adiabaticexpansionN}. Again, in order to find the desired asymptotic behavior for the trigonometric phase $\Theta_k$ it is enough to evaluate the adiabatic expansion at $\tau=0$. At this limit we obtain a well-defined large $k$ asymptotic expansion 
\be
h^{I}_k(0)\sim \frac{1}{\sqrt{2}}\Big(1-i\frac{ \gamma}{4 k^2}-\frac{\gamma^2}{32k^4}-i\frac{21 \gamma^3}{128 k^6}-\frac{85 \gamma^4}{2048 k^8}+\cdots \Big) \, ,
\ee
and $h_k^{II}(0)\sim  h_k^{I*}(0)$, which requires the following large $k$ expansion for $\Theta_k$: 
\be \label{adiabaticthetaf}\Theta_k \sim  -\Big(\frac{\gamma}{4k^2}+\frac{\gamma^3}{6 k^6}+\frac{4\gamma^5}{5 k^{10}} + \cdots \Big) \ .  \ee
This determines the appropriated rate for the decaying of $\Theta_k$ when $k\to \infty$ to have a $CPT$-invariant vacuum of infinite adiabatic order. A vacuum that satisfies the asymptotic condition above is an adiabatic vacuum state and hence ultraviolet regular or,  equivalently, Hadamard.

\section{States of Low Energy for fermions}\label{Sec::SLE::Fermions}

In this section, we extend the prescription to build States of Low Energy to spin-$\frac{1}{2}$ fields. We proceed here in analogy with the scalar field case. First, we consider a generic scale factor $a$, and then, we particularize the method for a radiation-dominated universe with CPT symmetry.  Although we do not present here a formal proof that the states of low energy in a general FLRW are Hadamard, we check that the resulting $CPT$-invariant states of low energy condidered here satisfy \eqref{adiabaticthetaf}, and therefore are Hadamard/adiabatic states.\\ %that we are considering through this paper, 
%are all Hadamard as stated by \eqref{adiabaticthetaf}.% We will leave the general proof for future work.}\\

The starting point is again to fix a basis of solutions for the modes, namely, $\{ h^{I}_k,h^{II}_k\}$. Any other set of modes can be parametrized in the form
\bea
t^I_k&=&\lambda_k \, h^{I}_k+\mu_k \, h^{II*}_k\, ,\\
t^{II}_k&=&\lambda_k \,h^{II}_k-\mu_k \, h^{I*}_k\, .\nonumber 
\eea
From the normalization condition, $\mu_k$ and $\lambda_k$ should obey
\be
|\lambda_k|^2+|\mu_k|^2=1\, .
\ee
 The smeared energy density over a temporal window function $f^2$ is given by (see Sec. \ref{sec::SLE::Scalars})
\be 
\mathcal{E}_k[f]:=\int \text{d}\tau \sqrt{|g|} \,f^2\, \rho_k\, .
\ee
where $\sqrt{|g|} =a^{4}$, and where the energy density $\rho_k$ associated with the set of modes $\{t^{I}_k,t^{II}_k\}$ reads %\cite{DNT}
\be
 \rho_k(\tau)=\frac{2 i}{a^4}\left( t_{k}^{I} \frac{\partial t_{k}^{I *}}{\partial \tau}+t_{k}^{I I} \frac{\partial t_{k}^{I I *}}{\partial \tau}\right)\, .\ee

We can choose $\mu_k$ or $\lambda_k$ to be real since $\left\{e^{i\alpha}\,t^{I}_k , e^{i\alpha}\, t^{II}_{k} \right\}$ is also a solution of the system of equations. For future convenience, and  following similar arguments than in the scalar case \cite{Olbermann07}, we assume that $\lambda_k$ is real and positive $\lambda_k>0$. %We  take $\lambda_k$  real for future convenience. 
The smeared energy density can be written, as in the scalar case, in terms of two constants $c_{1}\equiv c_{k,1}$ and $c_{2}\equiv c_{k,2}$, namely
\bea\label{Wkfermions}
 \mathcal{E}_k&=&(1-2|\mu_k|^2)c_{1}+2 |\mu_k|\ \,  \textrm{Re}(\mu^*_k c_2)\\
 &\equiv&(1-2|\mu_k|^2)c_{1} + 2 |\mu_k|\sqrt{1-|\mu_k|^2}\,|c_2|\cos(\textrm{Arg}c_2-\textrm{Arg}\mu_k)\, , \nonumber
\eea
where
\bea
c_{1}&=&2i\int \text{d}\tau  f^2\, \left( h_{k}^{I} \frac{\partial  h_{k}^{I* }}{\partial \tau}+h_{k}^{I I} \frac{\partial \ h_{k}^{I I *}}{\partial \tau}\right)\, ,\label{c1fermions}\\
c_2&=&2i\int \text{d}\tau  f^2\, \left( h_{k}^{I} \frac{\partial  h_{k}^{II}}{\partial \tau}-h_{k}^{I I} \frac{\partial  h_{k}^{I }}{\partial \tau}\right) \, . \label{c2fermions}
\eea
 From now on, we assume that the fiducial modes are such that  $c_1$ is a real, negative quantity. Note that this is the case for the standard mode solutions in Minkowski spacetime.\\
 
We find that $\mathcal{E}_k$ is trivially minimized with respect to $\textrm{Arg} \mu_k$ for $\textrm{Arg} \mu_k= \textrm{Arg}c_2+\pi$. Therefore, the task now is to minimize 
%\be \label{eq:minferm}
%\mathcal{E}_k=(2\lambda_k^2-1)c_{1} - 2 \lambda_k\sqrt{1-\lambda_k^2}\,|c_2|\, .
%\ee
\be \label{eq:minferm}
\mathcal{E}_k=(1-2|\mu_k|^2)c_{1} - 2 |\mu_k|\sqrt{1-|\mu_k|^2}\,|c_2|\, .
\ee
 with respect to $|\mu_k|$. Taking $\partial_{|\mu_k|}\mathcal{E}_k =0$ % $ \partial_{\lambda_k}\mathcal{E}_k =0$
 we obtain four possible solutions. Only two of them are real and positive for $\lambda_k$,
\be \label{eq:lambdaf}
\lambda_k=\sqrt{\frac{1}{2}\mp \frac{c_{1}}{2\sqrt{c_{1}^2+|c_2|^2}}} \,, \qquad |\mu_k|=\sqrt{\frac{1}{2} \pm \frac{c_{1}}{2\sqrt{c_{1}^2+|c_2|^2}}}\, .
\ee
 From the above solution, one can easily check that the one that minimizes the smeared energy density $\mathcal{E}_k$ is %Inserting \eqref{eq:lambdaf} into Eq. \eqref{eq:minferm} we find that the solution that minimizes $\mathcal{E}_k$ is 
\be \label{eq:lambdaf2}
\lambda_k=\sqrt{\frac{1}{2}- \frac{c_{1}}{2\sqrt{c_{1}^2+|c_2|^2}}} \,, \qquad |\mu_k|=\sqrt{\frac{1}{2} + \frac{c_{1}}{2\sqrt{c_{1}^2+|c_2|^2}}}\, .
\ee
and the minimum value of the smeared energy density $\mathcal{E}_k$ is $ \mathcal{E}_k= -\sqrt{c_1^2 + |c_2|^2}$.\\

As a final remark, we note that, in contrast to the analysis for scalar fields, it is here possible to find a maximal value for $\mathcal{E}_k$. If we take now $\textrm{Arg} \mu_k=\textrm{Arg}c_2$ in \eqref{Wkfermions} and then compute $\partial_{|\mu_k|}\mathcal{E}_k =0$ we find again \eqref{eq:lambdaf}. Now, inserting them into $\mathcal{E}_k$ again, we find that it takes its maximum value for the opposite solution (+ for $\lambda_k$ and $-$ for $\mu_k$), that gives $ \mathcal{E}_k=\sqrt{c_1^2 + |c_2|^2}$. This renders a nonphysical state.

\subsection{$CPT$-invariant States of Low Energy}
We can repeat the above analysis imposing CPT invariance. In this case, we can choose a convenient fiducial solution  given by 
\be \label{eq:fiducial:fermions} h_k^{I}=h_k^{I;CPT}(\tau,\Theta_k=0)\, , \qquad  h_k^{II}=h_k^{II;CPT}(\tau,\Theta_k=0)\, ,\ee
with $h^{I;CPT}_k$ and $h^{II;CPT}_k$ given in Eqs. \eqref{h1rad-cpt} and \eqref{h2rad-cpt} for $\Theta_k=0$. Therefore, the modes $t_{k}^{I,II}$ read
 \bea
t^I_k&=&\cos(\Theta_k) h^{I}_k\,\,+i\sin(\Theta_k)  h^{II*}_k\, ,\\
t^{II}_k&=&\cos(\Theta_k)  h^{II}_k-i\sin(\Theta_k)  h^{I*}_k\, .
\eea
Therefore, 
\be
\mathcal{E}_k=\cos(2 \Theta_k) c_{1} +\sin(2 \Theta_k) \textrm{Im}(c_2) \, ,
\ee
 and the minimization equation $\partial_{\Theta_k}\mathcal{E}_k=0$ becomes 
 \be
 -\sin(2 \Theta_k) c_{1}+\cos(2 \Theta_k) \textrm{Im}( c_2)=0\, ,
 \ee
 therefore
 \be \label{eq:thetaf-low}
 \tan(2 \Theta_k)=\frac{\textrm{Im}( c_2)}{c_{1}}\, .
 \ee
 
The solution for the angle then reads
\be \label{ff}
\Theta_k=\frac{1}{2}\arctan\left(\frac{\textrm{Im}(c_2)}{c_1}\right)+\frac{n \pi}{2}.
\ee
For $n$ even we have a state of low energy, while for $n$ odd we obtain a state of high energy, which has to be discarded. Up to irrelevant global phases (\ref{ff}) characterize a single low energy state, $CPT$-invariant, depending only on the smearing function $f^2$.
\subsubsection{An example: $CPT$-invariant vacuum of low energy at late times }

We can also compute the particle creation for an initial vacuum state characterized by $\Theta_k$. The vacuum $|0\rangle$ is perceived at late times as a collection of particles, defined as quantum excitations of the  adiabatic {\it out}-vacuum  $|0_{+}\rangle$. We find\footnote{Note that the spectrum is indeed independent of the helicity $h$.} 
\be
n_{k,h}= |\beta_{k, h}|^2=\frac{1}{2}-\frac{e^{-\pi \kappa}\sinh(2\pi \kappa)\sqrt{\kappa}}{4\pi}\Bigg(e^{- 2 i \Theta_k}e^{i\frac{\pi}{4}}\Gamma(i \kappa)\Gamma(\tfrac{1}{2}-i \kappa)+ e^{2 i  \Theta_k}e^{-i\frac{\pi}{4}}\Gamma(-i \kappa)\Gamma(\tfrac{1}{2}+i \kappa)\Bigg)\ , 
\ee
where $\kappa=\frac{k^2}{4 \gamma}$.  As for the case of scalar fields, and in agreement with the results of \cite{LetterCPT, Large}, the above expression can be rewritten as 
\be
|\beta_{k, h}|^2=\frac{1}{2}\big(1-\cos(2\eta_k)\cos(\Lambda_k)\big) \ , 
\ee
where $\cos(\Lambda_k)=\sqrt{1-e^{-4\pi \kappa}}$  and
\be \label{eq:phases-f}
2\eta_k=2(\Theta_k^{late}-\Theta_k) \, 
\ee
with
\be \label{cptvacuumf}\Theta_k^{late} = \frac{\pi}{8}+\frac{1}{2} \textrm{Arg}[\Gamma(\tfrac{1}{2}-i\kappa)\Gamma(i\kappa)] \, 
\ . \ee
%\be2\eta_k=-2\theta_k^f +\frac{\pi}{4}+ \textrm{Arg}[\Gamma(\tfrac{1}{2}-i\kappa)\Gamma(i\kappa)]\ . \ee
Arg$(z)$ refers to the argument of $z$. As for the scalar case this state is the low energy state associated to a smearing function $f^2$ with support at $|\tau| \sim \infty$. It  minimizes the smeared energy density $\mathcal{E}_k$ at late times independently of $f^2$, and as expected, it is Hadamard. One can easily check this statement by evaluating the asymptotic large $k$ expansion of $\Theta_k^{late}$ and confirming that it agrees with the adiabatic expansion (\ref{adiabaticthetaf}) at any order.  This Hadamard  and $CPT$-invariant state turns out to be equivalent to the one proposed in \cite{LetterCPT, Large}.

\section{$CPT$-invariant States of Low Energy at $\tau=0$ for fermions }\label{Sec::SLE::Fer::Early}
In this section, we study how to obtain a  vacuum state with the SLE prescription using a smearing function with support around $\tau=0$. For this purposes, we  use again the Gaussian function $f_g^2$ defined in Eq.\eqref{Gaussian}. It is interesting to note that the energy density decays as $a^{-4}$ for $\tau \to 0$, this is a consequence that massless fermions enjoys conformal invariance. A similar behavior was found for massive scalars with $\xi=1/6$. It means that the term $\sqrt{|g|}=a^{4}$ from the volume element makes the integral of the smeared energy density perfectly finite (see also the conformally coupled scalar case Appendix \ref{App::SLE::Conf}). As in the scalar case, we study both the massless and the massive cases.

\subsection{Massless case}
In this case the fiducial solution \eqref{eq:fiducial:fermions} correspond to the conformal modes $h^{I}_{k}= \frac{1}{\sqrt{2}}e^{-i k \tau}$ and $h^{II}_{k}= \frac{1}{\sqrt{2}}e^{-i k \tau}$. For the smearing function we choose $f^2=f_g^2$. If we compute the integrals \eqref{c1fermions} and \eqref{c2fermions} we obtain

\be \label{eq:c1_c2_fermions_massless}
c_{1} =-\frac{2k}{2} \int \text{d} \tau\, f_g^2  = -\frac{k}{2} \, ,  \qquad
c_2 =0  \ .
\ee
Therefore from \eqref{eq:thetaf-low} we get the state of low energy \be \Theta^{m=0}_k=0\, .\ee This result  is independent of the smearing function that we use. This is because $\Theta_k=0$ minimizes the energy density
\be
\rho_k(\tau) = -\frac{2 k}{a^{4}}\cos{2\Theta_k} \, ,
\ee
for all $\tau$. We note that $\Theta_k^{m=0}=0$ satisfies the adiabatic condition \eqref{adiabaticthetaf} at all orders (remember that $m=0$ implies $\gamma=0$). This is the same situation as for the conformally coupled massless scalar field (see Appendix \ref{App::SLE::Conf}).  

\subsection{Massive case}

%We take again the fiducial solution given in \eqref{eq:fiducial:fermions}, and $f^2=f_g^2$ for the smearing function. As for the massive scalar case w
Let us study now the massive case. We follow the same procedure as in the scalar case to obtain an approximated state of low energy by expanding the modes $h^{I}_k$ and $h^{II}_k$ around $\tau\sim 0$.  We explicitly compute its large-$k$ behavior and check that it satisfies the adiabatic condition \eqref{adiabaticthetaf} up to a given order, that increases as we improve the orders of the expansion in $\tau$.\\

The process is as follows. First, we expand the modes in $c_1$ and $c_2$ in powers of $\tau$ around $\tau=0$. We use the fiducial solution given in \eqref{eq:fiducial:fermions} and the Gaussian smearing function $f^2=f_g^2$. The first orders of the expansion read
\bea
c_{1} = &&2i \int \text{d}\tau  f_g^2\,  \left( h_{k}^{I} \frac{\partial  h_{k}^{I*}}{\partial \tau}+h_{k}^{I I} \frac{\partial  h_{k}^{II*}}{\partial \tau}\right) = \int \text{d}\tau  f_g^2\left(-4 \sqrt{\gamma  \kappa } -\frac{2}{3} \gamma ^2\sqrt{\gamma  \kappa } \tau ^4  +\cdots\right) ,\label{c1fermionsexpan}\\
c_2=&& 2i \int \text{d}\tau  f_g^2\,\left( h_{k}^{I} \frac{\partial  h_{k}^{II}}{\partial \tau}-h_{k}^{I I} \frac{\partial  h_{k}^{I }}{\partial \tau}\right) \,= i \int \text{d}\tau  f_g^2 \left(4  \sqrt{\gamma ^3 \kappa }\tau ^2 -\frac{16}{3}  \gamma ^{5/2} \kappa ^{3/2} \tau ^4 +\cdots \right)\, . \label{c2fermionsexpan}
\eea
Then, we insert this integrals in \eqref{eq:thetaf-low} and obtain an approximated solution for the initial phase $\Theta_k$. We have computed these expressions up to $O(\tau^{14})$. Finally, we compute the large-$k$ expansion of $\Theta_k$, obtaining
\be
  \Theta_k =\frac12 \arctan\left(\frac{\textrm{Im}( c_2)}{c_{1}}\right)\sim -\Big(\frac{\gamma}{4k^2}+\frac{\gamma^3}{6 k^6}+\frac{4\gamma^5}{5 k^{10}} + O\left( k^{-14}\right)\Big)\, ,
\ee
which fully agrees with the asymptotic expansion  \eqref{adiabaticthetaf}. As for the scalar case, these computations have also required the assistance of the \emph{Mathematica} software. The more orders we consider in the $\tau$ expansion the better is the coincidence with the adiabatic expansion.  We have checked that for an expansion at order $O(\tau^{14})$ we recover the adiabatic expansion to $O\left( k^{-14}\right)$. This gives strong evidence that the prescription for the low energy states proposed here is consistent with the Hadamard/adiabatic condition for more general FLRW spacetimes. % {\color{red} revisar terme $\pi/4$. Aquest sembla que apareix perquè quan fem la expansió en tau }

\section{Conclusions and final comments}\label{sec:conclusions}

%{\color{red}[en els final comments podriem posar molt esquematicament algunes de les coses que va explicar Pepe en la xarrada de Genova sobre la trace anomaly]}
The concept of states of low energy appears to be a very useful prescription to single out a preferred state in FLRW cosmologies. One of the major virtues of the construction is that it guarantees the Hadamard condition for the selected vacuum state. The crucial point is the use of a smearing window in the time variable.  The prescription was established in \cite{Olbermann07} for scalar fields. In this paper, we have extended the construction to spin-$\frac{1}{2}$ fields and applied it to the special case of a radiation-dominated universe. In this context a further symmetry condition can also be imposed. In conformal time $\tau$, the expansion factor for a radiation-dominated universe is a linear function, which allows analytic continuation to negative values of the conformal time \cite{LetterCPT,Large}. %Time-reversal 
CPT symmetry at the big bang can be naturally required as an extra condition to impose on the low energy states. A possible choice for the smearing function is to select it with support at $|\tau|\to \infty$. %, as implicitly assumed in Refs. \cite{LetterCPT,Large}. 
In this case, the resulting state is independent of the particular choice of the smearing function, since its support lies in an adiabatic region of the spacetime.  However, this involves giving initial conditions by knowing the late-times behavior of the expanding universe. There is a more natural option that consists of choosing the window function around the big bang itself.  Performing a careful analysis for the minimization of the smeared energy density, including the appropriate  factors coming from the volume element, we have checked that this choice is fully consistent with physical requirements at the  ultraviolet, namely, the adiabatic/Hadamard condition. Therefore, these states are then suitable candidates as effective big bang vacua from the quantum field theory viewpoint. The infrared behavior of these states is then sensitive to the smearing function chosen. This ambiguity  %included in the choice of smearing function 
could be naturally interpreted, at least heuristically, as encoding quantum gravity effects of a more fundamental theory. \\

\section*{Acknowledgments}

 We thank the anonymous referee for their careful reading of this manuscript and very
helpful comments. We have benefited from discussions with I. Agulló, A. Ferreiro, and J. Olmedo. We also thank P. Beltrán-Palau for collaborating in early stages of this project. This work is supported by the Spanish Grants No. %FIS2017-84440-C2-1-P funded by MCIN/AEI/10.13039/501100011033 “ERDF A way of making Europe”, Grant
 PID2020-116567GB-C2-1  funded by MCIN/AEI/10.13039/501100011033, 
 and  PROMETEO/2020/079 (Generalitat Valenciana). 
S. N. is supported by the Universidad de Valencia, within the Atracci\'o de Talent Ph.D Fellowship No. UV-INV- 506 PREDOC19F1-1005367. S. P. is supported by the Leverhulme Trust, Grant No. RPG-2021-299. %, and was previously supported by the Ministerio de Ciencia, Innovaci\'on y Universidades, Ph.D. fellowship, Grant No. FPU16/05287. \\

\appendix

\section{The adiabatic expansion} \label{ap:adiabatic}

%In the following 
In this appendix we briefly review the adiabatic method for scalar fields. We follow \cite{parker-toms}, but we translate the notation and expression to  work in conformal time. Consider a massive scalar field $\phi$ propagating in a flat FLRW spacetime $\text{d}s^2=a^2(\text{d}\tau^2-\text{d}\vec x ^2)$. As in the main text, we expand the quantized field in Fourier modes \eqref{modecomp}. From the Klein-Gordon equation, we can easily obtain the equation for the field modes
\be
\phi_k^{\prime \prime}+2 \frac{a^{\prime}}{a} \phi_k^{\prime}+\Big(k^{2}+a^{2}m^2+6 \xi \frac{a^{\prime \prime}}{a}\Big) \phi_{k}=0\, .
\ee
As in the main text, it is convenient  to work with the rescaled Weyl field $\varphi \equiv a \phi$ and the rescaled  modes $\varphi_k (\tau)\equiv a(\tau) \phi_k(\tau)$. And then, the mode equation results in
\be\label{eq:ap:mode-varphi}
\varphi''_k +\Big(\omega^2+(6\xi-1)\frac{a''}{a}\Big)\varphi_k=0\, .
\ee

An unavoidable requirement that any suitable vacuum state must meet is that is has to be ultraviolet regular. It can be easily understood by requiring that the short distance behavior of the Feynman Green's function $iG_F(x,x')$ and related quantities must be similar to that found in Minkowski space. This becomes necessary to guarantee the existence of finite vacuum expectation values after renormalization.  For quantum states in FLRW spacetimes this criterion can be implemented by the {\it adiabatic condition} \cite{parker-toms} [Sec. 3.1]. In terms of field modes this means that, for large $k$, 
 the field modes must behave as
\be \label{eq:ap:conformalmodes}\varphi_k(\tau)\sim \frac{1}{\sqrt{\Omega_k(\tau)}}e^{-i\int^\tau \Omega_k(\tau') d\tau'}  \ , \ee 
where the function $\Omega_k(\tau)$ admits an asymptotic adiabatic expansion in terms of the derivatives of $a(\tau)$
\be \label{adiabaticexpansionN-ap}\Omega_k = \omega_k^{(0)} + \omega_k^{(1)}+ \omega_k^{(2)} + \omega_k^{(3)}+ \omega_k^{(4)} + \cdots   \ . \ee
The coefficient $\omega_k^{(n)}$ depends on derivatives of $a(\tau)$ up to and including the order $n$.  The leading order of the expansion is $\omega_k^{(0)}\equiv \omega=\sqrt{k^2+a^2 m^2}$ and the next-to-leading orders are obtained, by systematic iteration, from the relation
\be \label{ad_eq}
\Omega_k^2=\omega^2+(6\xi-1)\frac{a''}{a} + \frac{3}{4}\frac{ (\Omega'_k)^2}{\Omega^2_k}-\frac{1}{2}\frac{ \Omega''_k}{\Omega_k} \ ,
\ee
derived from the mode equation \eqref{eq:ap:mode-varphi}. Inserting the adiabatic expansion in the equation above, and grouping terms with the same adiabatic order, it is possible to obtain the $n$th coefficient from the lower ones once the leading term is defined. It can be proved that the terms with odd adiabatic order are zero, i.e., $\omega_k^{(2n+1)}=0$. The first next-to-leading order terms can be found, for example, in \cite{parker-toms}. From the adiabatic expansion of the field modes, one can easily build the adiabatic expansion of the Feynman Green's function at coincidence%\footnote{{\color{blue}\sout{We remark that this expansion is equivalent to the DeWitt-Schwinger proper-time expansion expansion of the Feynman Green's function when evaluated in flat FLRW spacetimes. See Ref.} \cite{beltran-nadal1,rio-navarro}. \sout{The equivalence also holds at separated points.}}}
\be \label{eq:G:ad:coincidence}
i G_F(x,x)_{\textrm{Ad}}=\int\frac{\text{d}^3k}{2(2 \pi )^3a^2}\sum_{n=0}^{\infty}(\Omega_k^{-1})^{(n)}\, .
\ee
%{\color{blue}\sout{We note that this expansion, when expressed at separate points,  coincides with the deWitt-Schwinger expansion of the two-point function }\cite{beltran-nadal1,rio-navarro,Silvia-Winstanley},\sout{ and hence it satisfies the Hadamard condition (see also } \cite{Pirk93,Hollands01}).
In Refs. \cite{beltran-nadal1,rio-navarro,Silvia-Winstanley} it was checked that the first orders of this expansion, when expressed at separated points, coincide with the deWitt-Schwinger expansion of the two point function in four spacetime dimensions. As stated above, in FLRW universes the Hadamard condition translates to require for the field modes $\varphi_k$ a large momentum behavior dictated by \eqref{conformalmodes} at all orders. A state that satisfies this requirement is called a state of {\it infinite adiabatic order} (or just an adiabatic state). %In some references \cite{4adiabatic,agullo-nelson-ashtekar}, it was stressed that to obtain a finite value of the stress-energy tensor after renormalization is enough to ensure that the (divergent) large frequency behavior of the modes $\varphi_k$ coincides with the large frequency behavior of the adiabatic expansion up to $4th$ order.  However, this requisite is not strict enough since it is always possible to build some other composite operators (polynomials of the scalar field) that cannot be appropriately renormalized with this type of states \cite{Fewster13}. For this reason 
We demand the physical admissible states to be adiabatic states (of infinite adiabatic order) and hence equivalent to be Hadamard states. %{\color{blue} A phrase for fermions and references} \\

\section{Large $k$ expansion at a fixed time}\label{ap:expansions}

In this appendix, we give, for a radiation-dominated spacetime, the large-$k$ expansion of the $CPT$-invariant two-point function in  momentum space at a fixed time and for an arbitrary $\theta_k$, and compare it with its adiabatic expansion. Our goal here is to show that only if $\theta_k$ satisfies the asymptotic condition given in \eqref{aexpansiontheta}, the field modes are compatible with the adiabatic condition, and therefore, the vacuum state is Hadamard. \\

From the $CPT$-invariant solution \eqref{eq:sol-scalars-cpt}, and for an arbitrary value of $\theta_k$, we can compute the large $k$ expansion of the  square of the $CPT$-invariant modes at a fixed time $\tau$
\bea \label{eq:largek-scalars1}
|\varphi_k^{CPT}|^2 \sim && \,\cosh (2 \theta_k ) \left(\frac{1}{k}-\frac{\gamma ^2 \tau ^2}{2
   k^3}+\frac{3 \gamma ^4 \tau ^4}{8 k^5} +\frac{\gamma ^2}{4 k^5} -\frac{\gamma ^2
   \cos (2 k \tau )}{4 k^5}+  \mathcal{O}\left( k^{-6}\right)\right) \\
   \nonumber\\
   &&+ \sinh (2 \theta_k) \left(-\frac{\gamma ^2}{4 k^5} +\frac{\cos (2 k \tau )}{k}-\frac{\gamma
   ^2 \tau ^3 \sin (2 k \tau )}{3 k^2}-\frac{\gamma ^4 \tau ^6 \cos
   (2 k \tau )}{18 k^3}-\frac{\gamma ^2 \tau ^2 \cos (2 k \tau )}{2 k^3} +  \mathcal{O}\left( \frac{\sin(2k\tau)}{k^{4}}\right) \right)  \nonumber
\ , \eea
and compare this result with the asymptotic behavior dictated by the adiabatic expansion, namely

\be \label{eq:largek-scalars2}
|\varphi_k |^2_{\textrm{Ad}}\sim \frac{1}{k} -\frac{\gamma ^2 \tau ^2}{2 k^3} +\frac{\gamma ^2}{4 k^5}+\frac{3 \gamma ^4 \tau ^4}{8 k^5}+ \mathcal{O}\left(k^{-6}\right) \, .
\ee
If we now impose for the initial hyperbolic phase the asymptotic behavior given in \eqref{aexpansiontheta}, the oscillatory behavior in \eqref{eq:largek-scalars1} cancels out and we recover, order by order, the large $k$ behavior required by the adiabatic expansion \eqref{eq:largek-scalars2} at any time $\tau$. Therefore, any $\theta_k$ obeying \eqref{aexpansiontheta} gives an adiabatic (Hadamard) $CPT$-invariant vacua.\\

\section{States of Low Energy for conformally coupled scalars}\label{App::SLE::Conf}

 For conformally coupled scalar fields $\xi=1/6$ it is convenient to write the energy density $\rho_k$ for the mode $T_k$ in terms of the Weyl transformed mode $\mathcal{T}_k$ (i.e., $\mathcal{T}_k=a\,T_k$) because it takes the simple form
\be
\rho_k(\tau)=\frac{1}{4 a^4}\left(|\mathcal{T}'_k|^2+\omega^2 |\mathcal{T}_k|^2\right)\, ,
\ee
with $\omega^2=k^2 +m^2a^2$.  Note that all the divergent behavior at $\tau\to 0$ is encapsulated in the term $a^{-4}$. The minimization prescription follows as in Sec. \ref{sec::SLE::Scalars}. The smeared energy density to be minimized around the big bang takes the simple form
\be
\mathcal{E}_k[f] :=\int \text{d} \tau\, \sqrt{|g|}\, f^2\, \rho_k = \int \text{d} \tau\,  f^2\, \frac{1}{4}\left(|\mathcal{T}'_k|^2+\omega^2 |\mathcal{T}_k|^2\right) \, ,
\ee
As we see in the above equation, conformally coupled scalar fields are less sensitive to the big bang singularity. The same behavior was found for spin-$\frac12$ fields. We try to define a Hadamard state around $\tau=0$ as we did in the above sections. The integrals which are left to compute are

\bea \label{eq:ap:c1_c2_conformal_Weyl}
c_{1}&\equiv&c_{k,1}=\frac{1}{4} \int \text{d} \tau f^2\,\left(|\varphi'_k|^2+\omega^2 |\varphi_k|^2\right),\\c_2&\equiv&c_{k,2}=\frac{1}{4} \int  \text{d} \tau f^2\, \left(\varphi'{}_k^2+ \omega^2 \varphi_k^2 \right)\, .
\eea
where $\left\{\varphi_k, \varphi_k^*\right\}$ are a pair of fiducial solutions of the Klein-Gordon equation. Any general mode is given by
\be
\mathcal{T}_k(\tau) = \lambda_k \varphi _k (\tau)+\mu_k \varphi_k^*(\tau)
\ee
For the massless case we can take the conformal vacuum $\varphi_k(\tau)=\frac{e^{-ik\tau}}{\sqrt{k}}$ as the fiducial mode and obtain that $c_2=0$ irrespectively of $f^2$. Therefore, using \eqref{mu-lambda}, we find $\mu_k=0$ and $\lambda_k=1$, and we conclude that the conformal vacuum minimizes the energy density at any $\tau$. Therefore, this state is independent of the test function.  One can also see this by computing the energy density for any $CPT$-invariant state parametrized by $\theta_k$. In this case one obtains $\rho_k= \frac{k}{{\color{red}4} a^4} \cosh(2 \theta_k)$. This quantity is minimized for $\theta_k =0$, at any $\tau$. \\

For the massive case we can take Eq. \eqref{eq:basis-CPT-scalars} %$\varphi_k(\tau)=\varphi^{CPT}_k(\tau,\theta_k=0)$ 
as the fiducial solution and build a state that minimizes the smeared energy density with $f^2$ with support at $\tau=0$. As for spin-$\frac{1}{2}$ fields, we do not need extra requirements for the smearing function. The resulting Hadamard state is also dependent on $f^2$. 

\section{States of Low Energy in Minkowski for fermions}
\label{ap:MINKOWSKI}
Let us consider the minimization prescription described in Sec. \ref{Sec::SLE::Fermions} for fermions in a Minkowski spacetime. A general mode $\{t^I_k,t^{II}_k\}$ solution can be described by the following linear combination of fiducial solutions
\bea
t^I_k&=&\lambda_k \, h^{I}_k+\mu_k \, h^{II*}_k\, ,\label{t1t2Ap}\\
t^{II}_k&=&\lambda_k \,h^{II}_k-\mu_k \, h^{I*}_k\, .\nonumber 
\eea
We take the well known postive and negative-frequency solutions of Minkowski spacetime as the fiducial solutions.
\be\label{hIhIIMink}
h^{I}_k=  \sqrt{\frac{\omega + m  }{2 \omega}} e^{-i\omega \tau} \, , \qquad h^{II}_k=\sqrt{\frac{\omega -m }{2 \omega}} e^{-i \omega\tau}\, .
\ee
By minimizing the smeared energy density for a generic mode we obtain that only  $\lambda_k =1$ and $\mu_k=0$  renders a minimum value for the smeared energy density $\mathcal{E}_k$, irrespective of the test function used. This solution as one can see from \eqref{t1t2Ap} corresponds to the standard positive frequency mode in Minkowski spacetime. The calculation proceeds as follows. First one has to compute the coeficients $c_1$ and $c_2$ given in \eqref{c1fermions} and \eqref{c2fermions}. By substituting with \eqref{hIhIIMink} one arrives at
\be
c_{1}= -2\omega \int \text{d}\tau\, f^2\, \qquad c_2=0\, .
\ee
Therefore, the two possible solutions that appears when minimizing $\mathcal{E}_k$ [see Eq. \eqref{eq:minferm}] are given by
\be
\lambda_k=\sqrt{\frac{1}{2}\mp\frac{-\omega}{2\omega}}\, , \qquad|\mu_k|=\sqrt{\frac{1}{2}\pm\frac{-\omega}{2\omega}}\, .
\ee
From the above solutions, the one that minimizes the smeared energy density is given by
\be
\lambda_k=1 \, , \qquad \mu_k=0\, , \label{ap:eq:first:mink}
\ee
which corresponds to the standard positive-frequency solution \eqref{hIhIIMink}. If we compute the energy density we obtain 
\be
\rho_k=2 i\left( h_{k}^{I} \frac{\partial h_{k}^{I *}}{\partial \tau}+h_{k}^{I I} \frac{\partial h_{k}^{I I *}}{\partial \tau}\right)=-2\omega\, .
\ee
In other words, this choice gives the well-know state of low energy in Minkowski spacetime. This is the state of lowest energy for fermions. As a curiosity, the other possible solution that we obtain when finding the extrems of $\mathcal{E}_k$ corresponds to 
%which corresponds to the positive frequency solution from which we define the standard vacuum in Minkowski such that we have positive energies for the one particle/antiparticle state $H|k,s\rangle =\omega_k|k,s\rangle $. While the second solution 
$\lambda_k=0$, $\mu_k=1$.  %so thatt^{I}_k=-h_k^{II*}\, , \qquad t^{II}_k=h_k^{I*}\, , \ee
That correspond to negative frequencies, which renders a nonphysical state. In this case, the vacuum energy density results in $\rho_k=+2 \omega$,
%\be\rho_k=2 i\left( h_{k}^{II*} \frac{\partial h_{k}^{II}}{\partial \tau}+h_{k}^{I*} \frac{\partial h_{k}^{I}}{\partial \tau}\right)= +2\omega\, ,\ee
which corresponds to its maximal value. %the maximum value that the energy can take. % We note that this choice is forbidden since this corresponds to associating negative-frequency solutions to $B_{\vec{k} \lambda}$ in \eqref{psiexpansion}. This would produces negative energies for the one particle/antiparticle state $H|k,s\rangle =-\omega_k|k,s\rangle $.

%Let us take a closer look to the energy density:\be \rho_k(\tau)= i\left( t_{k}^{I} \frac{\partial t_{k}^{I *}}{\partial \tau}+t_{k}^{I I} \frac{\partial t_{k}^{I I *}}{\partial \tau}\right)- i\left( t_{k}^{I*} \frac{\partial t_{k}^{I }}{\partial \tau}+t_{k}^{I I*} \frac{\partial t_{k}^{II}}{\partial \tau}\right) \equiv 2 i\left( t_{k}^{I} \frac{\partial t_{k}^{I *}}{\partial \tau}+t_{k}^{I I} \frac{\partial t_{k}^{I I *}}{\partial \tau}\right)\, ,\eewhere we have used the derivative of the Wronskian condition to simplify the expression. 

\bibliography{bibliography}

\end{document}